\documentclass{ar-1col-S2O}
\usepackage[numbers]{natbib}
\usepackage{url}
\usepackage[utf8]{inputenc}
\usepackage{textcomp}
\usepackage[hidelinks]{hyperref} 

\begin{document}

\markboth{Gorgulla C.}{Virtual Screening Approaches}

\title{Recent Developments in Structure-Based Virtual Screening Approaches}

\author{Christoph Gorgulla$^{1,2,3}$
\affil{$^1$ Harvard Medical School, Harvard University, Boston, Massachusetts 02130, USA; email: c.gorgulla@gmail.com}
\affil{$^2$ Physics Department, Harvard University, Cambridge, Massachusetts 02138, USA}
\affil{$^3$ Department of Cancer Biology, Dana Farber Cancer Institute, Boston, Massachusetts 02215, USA}}

\begin{abstract}
Drug development is a wide scientific field that faces many challenges these days. Among them are extremely high development costs, long development times, as well as a low number of new drugs that are approved each year. To solve these problems, new and innovate technologies are needed that make the drug discovery process of small-molecules more time and cost-efficient, and which allow to target previously undruggable target classes such as protein-protein interactions. Structure-based virtual screenings have become a leading contender in this context. In this review, we give an introduction to the foundations of structure-based virtual screenings, and survey their progress in the past few years. We outline key principles, recent success stories, new methods, available software, and promising future research directions. Virtual screenings have an enormous potential for the development of new small-molecule drugs, and are already starting to transform early-stage drug discovery.
\end{abstract}

\maketitle

\tableofcontents

\section{INTRODUCTION}

The development of novel small-molecule drugs is a highly challenging process, which is time-consuming and expensive \cite{DiMasi2016a}. One critical step in this process is the discovery and optimization of hit and lead compounds. In the past few decades, experimental high-throughput screening (HTS) has been the primary method of discovering new hit compounds, and this approach is still the dominant method today. However, HTSs often suffer from several major disadvantages and challenges:


\begin{itemize}
    \item \textbf{Time.} HTSs can be time-consuming, e.g. because the binding assays that will be used in the screen have to be prepared and optimized. The ligand library that will be screened has to be prepared as well. The HTS itself can also be time-consuming, depending on the binding assays used and the equipment that is available to carry out the screen. 
    \item \textbf{Costs.} HTSs tend to be quite expensive, not least because of the robots that are needed to carry out the screens, as well as the physical ligand library \cite{Martis2011}.
    \item \textbf{Potency.} Initial hit compounds obtained by HTSs are often only moderately biologically active, which can in most cases be attributed to the fact that the screened libraries are of relatively small size (typically tens to hundreds of thousands of ligands).
    \item \textbf{Challenging targets.} As hits obtained by HTSs usually do not lead to very high-affinity binders, challenging target sites, such as flat protein-protein interactions or allosteric sites, can often not be successfully targeted. 
    \item \textbf{Scaffolds.} In HTSs that lead to sufficiently good hit compounds, the number of obtained hits and thus scaffolds are often quite limited. Furthermore, the same screening libraries (and thus scaffolds) are often used again and again in HTSs for different drug development campaigns, strongly limiting the chemical space of molecules that is explored for discovering new hit compounds.
    \item \textbf{Specificity} HTSs typically do not take into account whether the discovered hits are specific to the targeted receptor or are possibly binding to other biological macromolecules as well.
    \item \textbf{Binding mechanism.} HTSs do not provide mechanistic insight into how the ligands are binding to their target receptors. In phenotypic kind of screens, it is typically not even known to which receptor the ligand is binding.
    \begin{marginnote}[]
        \entry{Hit compound}{A molecule identified in a (virtual or physical) screening procedure that exhibits a sufficiently strong activity.}
        \entry{Lead compound}{A hit compound that was selected for further optimization regarding certain properties, such as potency, specificity, druglikeness, pharmacokinetic and pharmacodynamic properties.}
        \entry{High-throughput screening (HTS)}{Experimental technique in drug discovery using robotics to test thousands of molecules for their biological activity.}
    \end{marginnote}
\end{itemize}

Virtual screening approaches can mitigate all of the above problems, each to a different extent. Virtual screenings can be very fast and cost-effective. They can lead to highly potent initial hit compounds, and address even challenging target sites such as protein-protein interactions. In structure-based virtual screenings (SBVSs), the binding mode of the ligand to the target receptor is predicted, and thus (predicted) mechanistic insights are available. Virtual screens can also lead to larger number of hit compounds with different scaffolds, thus providing a range of backup compounds. 

\begin{marginnote}[]
    \entry{Structure-based virtual screenings (SBVS)}{Screens that are based on molecular dockings which require the structure of the target receptor.}
    \entry{Synthon}{Fragment together with reactivity information on how it can be used with other fragments to synthesize complete molecules.}
    \entry{Ligand-based virtual screens (LBVS)}{Screens that use the binding information of other ligands, but use no receptor structure.}
\end{marginnote}

\begin{figure}[t!]
    \includegraphics[width=0.75\paperwidth]{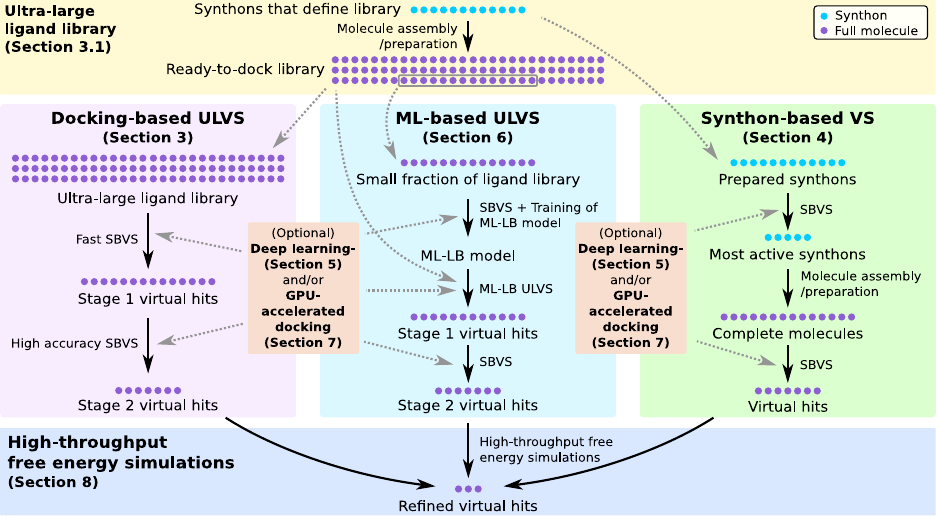}
    \caption{Overview of the major topics covered in this review, and how different techniques can be used in concert within the context of structure-based virtual screening approaches. Ultra-large ligand libraries (Section \ref{Sec:ULLL} are the starting point of several modern screening paradigms, such as docking-based ultra-large virtual screenings (Section \ref{Sec:ULVS}), synthon-based approaches (Section \ref{Sec:Synthon}), and ML-based ULVS (Section \ref{Sec:MLVS}). Deep learning-based (Section \ref{Sec:DL}) and GPU-based docking programs (Section \ref{Sec:GPU}) can be used to speed up structure-based virtual screens in any of the three primary virtual screening approaches mentioned above. Docking-based ULVSs require an ultra-large ligand library in a ready-to-dock format, while ML- and ligand-based ULVS (ML-LB-ULVS) require typically a SMILES version of the entire library and only a small fraction in ready-to-dock format (for training the ML model). Synthon-based approaches use the initial synthons that define the on-demand libraries for docking, and only require a small additional fraction of the assembled molecules in ready-to-dock format. High-throughput free energy simulations can be used as an additional screening stage to refine the results of all three primary virtual screening approaches discussed in this review.}
    \label{fig1}
\end{figure}

In this review, we first give an introduction to some of the fundamentals of virtual screenings (Section \ref{Sec:Fundamental}). Afterwards, we give an overview of recent developments regarding virtual screening approaches for drug discovery, starting with ultra-large virtual screens (ULVS, Section \ref{Sec:ULVS}). Ultra-large virtual screens in which more than 100 million ligands are screened. The subsequent sections cover approaches that can speed up ultra-large virtual screenings, and make them more cost-efficient. Among these approaches are synthon-based virtual screens (Section \ref{Sec:Synthon}), deep learning (DL) approaches to molecular docking (Section \ref{Sec:DL}), machine learning (ML) approaches to virtual screens (Section \ref{Sec:MLVS}) that combine structure-based virtual screens (SBVSs) with ligand-based ultra-large virtual screens (ML-LB-ULVSs), and hardware acceleration via GPUs (Section \ref{Sec:GPU}). Thereafter, we look into an approach that can increase the accuracy of virtual screenings (as well as their computational costs), namely their coupling with high-throughput free energy simulations (HTFES, Section \ref{Sec:HTFES}). Finally, we briefly discuss the potential that virtual screening approaches have in the future (Section \ref{Sec:Outlook}). A broad overview of the topics discussed in this review, and how they relate to each other, is shown in Figure \ref{fig1}. 

\section{FUNDAMENTALS}
\label{Sec:Fundamental}

Several concepts and underlying methods are used in connection with SBVSs. In this section, we introduce several of them. 

\subsection{Virtual Screenings}
Virtual screenings are procedures, in which a collection of ligands is computationally screened for their ability to bind to a given receptor structure. Receptors are most often proteins in the biomedical sciences but can be any type of biological macromolecule, including RNA or DNA. Virtual screens can be classified as ligand-based or structure-based. In ligand-based virtual screens, only the structure of known active compounds is used for the screening procedure to identify new active compounds. If the 3D structure of the receptor is available, structure-based methods can be used to carry out binding free energy calculations to estimate how strong the ligand binds to the targeted receptor, and the binding mode can be predicted  as well. There are a large number of free energy methods, and because of their high speed, the predominant type used in virtual screens are molecular docking procedures. Structure-based virtual screenings are normally preferred over ligand-based screens when the goal is to identify novel scaffolds, since ligand-based approaches are often biased towards compounds similar to known binders and do not provide a docking pose.

Virtual screens can be carried out in multiple stages, where in each subsequent step the best X compounds are used from the previous step, and methods of higher accuracy and/or computational are used to improve the quality of the dockings \cite{Gorgulla2020a,Olivet2022}. This principle is illustrated in Figure \ref{fig1}, where each stage acts as a funnel.

\subsection{Molecular Dockings}

A large variety of docking programs have been developed in the past several decades \cite{Biesiada2011,Fan2019,Pagadala2017,Rezacova2008b}. Docking programs typically consist of two components. First, a conformational search algorithm that explores the possible binding modes (conformation) of the ligand and the protein. The second part is the scoring function, which assigns to each conformation a docking score that correlates with the predicted binding affinity. The binding mode with the best score is typically the winning and final docking pose.

Docking algorithms can be classified into four primary categories: Physics-based (also called force field-based), knowledge-based (also called potential of mean force-based), empirical (also called regression-based), and machine learning-based (also called descriptor-based) docking routines that vary in the methods they utilize \cite{liu2015classification,yang2022protein,ain2015machine,li2019overview}. In Section \ref{Sec:DL}, we explore deep learning-based docking methods, which are a subclass of machine learning-based methods that have been developed only recently and can greatly speed up the docking procedures. 

\subsection{Receptor Structures and Preparation}

Structure-based docking requires the receptor structure as input. One primary source for receptor structures is the Protein Data Bank (PDB, \url{http://www.wwpdb.org/}), which is the largest database regarding experimentally determined protein structures \cite{Berman2003}. Here, X-Ray crystallography, nuclear magnetic resonance (NMR), as well as cryo-electron microscopy (cryo-EM) are the primary experimental methods to determine the structures of biological macromolecules. Cryo-EM is the youngest of these techniques and has experienced tremendous progress regarding its capabilities and resolution. Recently, even atomic resolution has been achieved \cite{nakane2020single}. Cryo-EM is particularly powerful regarding the determination of larger protein complexes, and thus also for the identification of protein-protein interactions. 

\begin{marginnote}[]
    \entry{Linear regression}{Method for predicting the relationship between a dependent variable and one or more independent variables via a (multi)linear function.}
    \entry{Knowledge-based method}{A method in the field of artificial intelligence that makes decisions based on the knowledge of a human experts stored in a knowledge base.}
    \entry{Physics-based method}{Method that use molecular force fields and the laws of physics directly in the computations, such as classical mechanics.}
\end{marginnote}

An alternative to experimentally determined structures is predicted structures. Here, homology modeling and \textit{de novo} structure prediction are the two primary types of prediction methods. Homology modeling uses a structure of a homologous protein that is known as a template. \textit{De novo} structure prediction does not require such a template and can predict new structures from scratch. Structure prediction, similar to cryo-EM, has experienced a breakthrough in the past few years. With the advent of AlphaFold \cite{Jumper2021,Jumper2022}, the structure of almost every protein known has been predicted and made freely available in the AlphaFold Structure Database \cite{Tunyasuvunakool2021,Varadi2022}. The AlphaFold Structure Database contains over 200 million protein structures. These recent developments in structure determination enable structure-based virtual screenings for almost any target protein.

\subsection{Ligand Preparation and Ligand Libraries}

Virtual screenings require ligand libraries that are used for the screening procedure. In the case of structure-based virtual screenings, the ligands have to be in a ready-to-dock format \cite{Gorgulla2020a}. Preparing ligands into a ready-to-dock format typically starts from the SMILES format and involves generating stereoisomers, tautomerization states, predicting protonation states, computing the 3D coordinates of the ligands, and converting them into the target formats. The precise formats that are needed depend on the docking programs that are used. Most docking programs require ligands in the PDBQT, PDB, MOL2, or SDF formats.

In ligand-based approaches, the ligands can be given in line notation, such as the SMILES format that is mostly used. However, sometimes 3D pharmacophore models are used in ligand-based screenings, in which case the ligands might need to be in a 3D format as well. 

\begin{marginnote}[]
    \entry{SMILES}{"Simplified Molecular Input Line Entry System”, is a line notation of molecules that encodes its topological structure.}
    \entry{PDBQT format}{An extension of the PDB (Protein Data Bank) format, used by the AutoDock family of docking programs, that also stores the partial charges (Q) and the AutoDock atom type (T).}
\end{marginnote}

\section{DOCKING-BASED ULTRA-LARGE VIRTUAL SCREENINGS}
\label{Sec:ULVS}

Ultra-large virtual screens have emerged as one of the most powerful virtual screening approaches to date. Virtual screenings have historically been called ultra-large when over ~100 million compounds are screened (though what ultra-large means might change in the future as computational power and library sizes increase. Several high-profile papers have been published in the last few years, including experimental validation, that have demonstrated the potential of this approach (see Section \ref{Ssec:history} for more details). Ultra-large virtual screenings have become possible in the past few years due to the recent advancements of multiple technologies and resources:
\begin{enumerate}
    \item \textbf{Ultra-large chemical libraries.} In the last decade, the first ultra-large ligand libraries with over 100 million commercially available molecules have become available.
    \item \textbf{Computational resources.} The capabilities of university computers clusters, national supercomputers, and commercial cloud computing infrastructures have improved substantially. Through these computational resources, sufficient computational power has become available to the scientific communities to carry out ultra-large virtual screenings. 
    \item \textbf{Software.} Software platforms, such as VirtualFlow \cite{Gorgulla2020a,Gorgulla2020b,Gorgulla2021b}, that can screen ultra-large ligand libraries using the above-mentioned computational resources, including cloud computing infrastructure, became freely available. 
\end{enumerate}

We will review available compound libraries and software in the sections below. 

\subsection{Ultra-Large Ligand Libraries}
\label{Sec:ULLL}

Ultra-large ligand libraries can be of different types. They can be \textit{public}, \textit{commercial}, or of \textit{proprietary} nature \cite{Hoffmann2019}. Here, \textit{public} refers to the free availability of the ligand database without general commercial availability, while commercial libraries contain compounds that are commercially available for purchasing. Proprietary libraries on the other hand are generally not available for purchasing, and usually only available for internal use by the company that possesses the library. Commercial libraries of ultra-large size contain mostly compounds that do not yet exist physically but can be synthesized efficiently on-demand, meaning upon ordering, by compound vendors with a certain success probability. These libraries are therefore also called on-demand libraries. Below we review ultra-large ligand libraries of public and commercial nature. Details on some proprietary libraries can be found in \cite{Hoffmann2019}. 

\subsubsection{Public Virtual Libraries}

Public virtual libraries can be useful and interesting because they allow to explore and test compounds via virtual screenings that are beyond current commercial availability, and with that vastly increase the chemical space that can be explored. Below we look at two types of these ultra-large public virtual libraries. 

\paragraph*{GDBs.} The Generated Databases (GDBs) by the Reymond Research Group likely were the first ligand libraries of ultra-large scale \cite{Meier2020,Reymond2012}. The GDB-11, containing all molecules with up to 11 atoms of type carbon, nitrogen, oxygen, and fluorine that satisfy certain valency, chemical stability, and synthetic accessibility criteria contained 110.9 million stereoisomers. Subsequently, additional GDB libraries were published. Among them are the GDB-13 containing molecules with up to 13 atoms resulting in 977 million molecules, as well as the GDB-17 with up to 17 atoms of the aforementioned atom types giving rise to 166 billion molecules \cite{Blum2009,Ruddigkeit2012}. While these libraries are among the first of ultra-large size, one disadvantage is that these libraries are only available in the SMILES format. Another disadvantage is that the compounds are not commercially available for direct purchasing. Rather, custom synthesis of these molecules is mostly required, which is possible via commercial companies but can be expensive and time-consuming. To address this problem, GDBChEMBL has been created, a library that contains only compounds from the GDB-17 that are similar to molecules in ChEMBL \cite{Buhlmann2020}. However, it is of relatively small size and only contains 10 million compounds. A similar library, GDBMedChem, was created that contains 10 million compounds from the GDB-17. It only contains compounds favorable from a medicinal chemistry point of view.

\paragraph*{KnowledgeSpace.} The KnowledgeSpace is another chemical space, which is similar to eXplore (see below), but more difficult to access in general \cite{Detering2010,BioSolveIT}. Like eXplore, it is based on commercially available building blocks and reactions known from the literature. However, there is no commercial service available that readily provides these compounds. With $2.9*10^{14}$ molecules, this space might be the largest public library for drug discovery available. It can be accessed via infiniSee from BioSolveIT \cite{BioSolveIT} but is not available in a ready-to-dock format.

\subsubsection{Commercial Libraries}

Ultra-large commercial libraries are based on a smaller number of fragments and combinatorial chemistry, which allows combining of the fragments via combinatorial chemistry to complete molecules. One important aspect regarding these on-demand libraries is their chemical diversity. Combinatorial libraries like these might lead to large numbers of compounds, but do the resulting compounds also represent a diverse chemical space, rather than a small region in the chemical space with highly similar compounds? Two recent studies have looked into this question. They concluded that the resulting chemical spaces are surprisingly diverse and more diverse than typical physical HTS libraries \cite{Tomberg2020,Irwin2020}. Below we review the major ultra-large on-demand libraries that are available.

\paragraph*{ZINC Libraries.} The two major disadvantages of the GDBs, non-commercial availability and non-ready-to-dock format were first addressed by the ZINC libraries \cite{Irwin2005,Irwin2012,Sterling2015,Irwin2020}. ZINC15 (published in the year 2015) was the first ultra-large ligand library of commercially available compounds in a ready-to-dock format, with initially approximately 120 million compounds \cite{Sterling2015}. The library has steadily grown since then and has reached approximately 1.5 billion molecules to date, out of which approximately 750 million are in a ready-to-dock format. A more recent version of the ZINC library is called ZINC20 (published in 2020) \cite{Irwin2020}. ZINC contains and aggregates the catalogs of hundreds of different compound vendors and other chemical libraries such as ChEMBL.

\paragraph*{REAL Database.} The REAL Database from Enamine was the first library available from a single chemical compound vendor reaching over 100 million compounds. The first version was published in 2007, containing approximately 29 million molecules \cite{Shivanyuk2007}. The REAL Database has grown over time, and the latest version of the REAL Database contains approximately 5.5 billion compounds \cite{EnamineREALDatabase} that comply with Lipinski's rule of five as well as Veber's rule. The majority of the compounds in the latest ZINC databases are from the REAL Database from Enamine. The typical time until shipping is 3-4 weeks, and the success rate of synthesis is approximately 80\%. 

\paragraph*{REAL Space.} The REAL Space, another on-demand library from Enamine, can be considered to be the larger sibling to the REAL Database. The REAL Space is mostly compliant with Lipinski's rule of five, but not completely \cite{Grygorenko2020}. This can be an advantage since there exist many drugs that are not rule of five compliant \cite{DeGoey2018}. The latest version of the REAL Space (year 2022) contains approximately 31 billion molecules \cite{EnamineREALSpace}. The REAL Space and the REAL Database overlap regarding the molecules they contain, with approximately 50-70\% of the compounds of the REAL Database being part of the REAL Space as well. In other words, approximately 10\% of the REAL Space is contained in the REAL Database.  The delivery time and synthesis success rate are roughly the same for the REAL Space and the REAL Database (approximately 80\%). 

\begin{marginnote}[]
    \entry{Lipinski's rule of five}{A rule of thumb that estimates whether a compound is druglike, specifically ($\textnormal{molecular weight}\le 500$, $\textnormal{logP}\le 5$, hydrogen bond acceptor count $\le 10$, and hydrogen bond donor count $\le 5$.}
    \entry{Veber's rule}{A rule of thumb that estimates whether a compound is druglike, specifically less than 10 rotatable bonds and a topological polar surface area of less than 140\AA$^2$.}
\end{marginnote}

\paragraph*{CHEMriya.} Another on-demand library, CHEMriya, is provided by Otava Chemicals \cite{CHEMriya}. The first version was released in the year 2021, and the latest version contains approximately 12 billion molecules. It is based on approximately 33,000 building blocks, and around 45 chemical reactions to combine these building blocks into complete drug-like molecules. CHEMriya can be accessed via infiniSee from BioSolveIT \cite{BioSolveIT}, but is not yet available in a ready-to-dock format.

\paragraph*{GalaXi.} WuXi AppTec also provides an on-demand chemical library, called GalaXi \cite{GalaXi}. The latest version of GalaXi contains approximately 8 billion molecules \cite{BioSolveIT}. The molecules can be synthesized within 4-8 weeks, and the success rate is approximately 60-80\% \cite{GalaXi}. Galaxi can be accessed via infiniSee from BioSolveIT \cite{BioSolveIT}, but is not available in a ready-to-dock format.

Regarding the major commercial on-demand libraries that are available, the REAL Database and REAL Space from Enamine, CHEMriya from Otava Chemicals, and the GalaXi Space from WuXi AppTec, one might wonder if and how much these libraries overlap in the chemical space. A recent study has looked at this aspect and found that the overlap is marginal between the REAL Space, CHEMriya, and the GalaXi Space \cite{Bellmann2022}. These three libraries contained at the point of study approximately 33 billion molecules jointly together. The number of molecules that were present in at least two of these libraries was less than 50 million, which is less than 1\% of the total number of molecules under consideration. 

\paragraph*{eXplore.} Another space that recently appeared is eXplore from eMolecules, containing 2.8 trillion molecules, being the largest on-demand chemical space for drug discovery to date \cite{eXplore,BioSolveIT}. Like the KnowledgeSpace, eXplore is based on building blocks that are commercially and readily available from various chemical compound vendors. In contrast to KnowledgeSpace, the synthesis steps that are used to combine the fragments into molecules are based on 40 robust reactions, mostly consisting of only one or two synthesis steps, facilitating the synthesis of the compounds. The final molecules can either be synthesized by the customers, or eMolecules can synthesize the molecules or for a fee. eXplore can be accessed via infiniSee from BioSolveIT \cite{BioSolveIT} but is not available in a ready-to-dock format.

\paragraph*{Freedom Space.} Another recent on-demand space is the Freedom Space from Chem-Space, located in Ukraine \cite{FreedomSpace,BioSolveIT}. This library is based on fragments from various compound vendors as well as known reactions. The library consists of 201 million compounds, 74\% of which comply with Lipinsiki's rule of five, and is based on fragments from various compound vendors. The synthesis is carried out by contract research organizations, and the delivery time is approximately 4 weeks. The Freedom Space can be explored via infiniSee from BioSolveIT \cite{BioSolveIT}, and can be downloaded in the SMILES format from the webpage of Chem-Space (\url{https://chem-space.com/compounds/freedom-space}).

\paragraph*{VirtualFlow Libraries.} The VirtualFlow open-source project currently provides two libraries in ready-to-dock format on the project homepage (\url{https://virtual-flow.org/}, \cite{Gorgulla2020a}). The first library is the REAL Database from Enamine (version of the year 2018), containing 1.4 billion molecules. The second library is ZINC library (year 2018), containing 1.4 billion molecules as well. 
In comparison to the original libraries, the VirtualFlow versions are completely in a ready-to-dock format and can be screened with VirtualFlow as well. In contrast, the original ZINC15/ZINC20 library so far only provides approximately 50\% of its molecules in a ready-to-dock format. And Enamine provides the REAL Database only in the SMILES format.

\subsection{History and Success Stories}
\label{Ssec:history}

Ultra-large virtual screenings are relatively new. Below we summarize some of the historical milestones, as well as major success stories. 

\paragraph*{Gorgulla (2018).} To our knowledge, the first ultra-large virtual screenings were reported in the dissertation of the author in the year 2018 \cite{Gorgulla2018}. In this thesis, VirtualFlow, an open-source virtual screening platform freely available to anyone was published (for more details see Section \ref{ssec:sw-vf}). VirtualFlow was applied in this work to prepare two snapshots of the ZINC library in a ready-to-dock format, one from the year 2014 containing 133 million ligands, and one from the year 2016 containing approximately 300 million ligands. The ultra-large virtual screens reported included a 130 million compounds screen against the target EBP1, a 180 million compound screen against the KIX-domain of MED15 to disrupt the MED15-SREBP interaction, and a 300 million compound screen targeting KEAP1 to disrupt the KEAP1-NRF2 interaction. In addition, this work provided a theoretical foundation on why the true hit rate is expected to increase with the size of the screened library. 

\paragraph*{Lyu et al. (2019).} In 2019, the first peer-reviewed ultra-large virtual screens were published by Brian Shoichet and John Irwin (the curator of the ZINC library) \cite{Lyu2019}, and contained extensive experimental validation. The first target in this study was the D4 dopamine receptor, a G-protein-coupled receptor (GPCR), that was screened with 138 million compounds. Picomolar compounds were discovered in this work directly from the virtual screen. The second target was AmpC $\beta$-lactamase, against which 99 million compounds were screened. Low-nanomolar binders were identified after optimization based on similarity searches of analog compounds in the same library, as well as their docking scores. This study, therefore, demonstrated that virtual screens and molecular docking cannot not only identify initial hits but also find improved analogs of existing hits. Another important result of this study is that it was experimentally observed that the true hit rate improves with the docking score, and thus the scale of the library that is screened. This result matches the theoretical prediction of this phenomenon in the author's dissertation \cite{Gorgulla2018}. 

\begin{marginnote}[]
    \entry{GPCR}{A large family of membrane proteins that convert extracellular signals into intracellular responses transduced primarily by G-proteins and $\beta$-arrestins.}
\end{marginnote}

\paragraph*{Stein et al. (2020).}
In this study, the first results of ligands originating from ultra-large virtual screenings involving animal studies are reported \cite{Stein2020}. Here, the Melatonin MT\textsubscript{1} receptor, which is a GPCR, was targeted with 150 million compounds from the ZINC library. The most potent identified ligand had a binding affinity in the picomolar range. Subsequent optimization lead to two inverse agonists, and mouse studies with these compounds showed that these compounds modulate the melatonin receptor biology in a target-specific way. 

\paragraph*{Gorgulla et al. (2020).}
In the year 2020, the VirtualFlow platform was published in Nature (see Section \ref{ssec:sw-vf} for more details), along with an experimental demonstration study targeting the KEAP1-NRF2 interaction \cite{Gorgulla2020a}. This study was the first to demonstrate that ultra-large virtual screenings can be used to target protein-protein interactions and the first study in which over 1 billion molecules were screened. The most potent compounds obtained were in the low nanomolar range and were able to disrupt the KEAP1-NRF2 interaction. In this study, a ready-to-dock version of the REAL Database from Enamine (version of the year 2018) was published, containing 1.4 billion molecules (see \ref{Sec:ULLL} for more details).

\paragraph*{Alon et al. (2021).} In this study, the $\sigma 2$ GPCR was targeted with ultra-large virtual screens \cite{Alon2021}. In total 31 compounds were obtained with submicromolar binding affinity. Three hits were optimized regarding their selectivity and their binding affinity and reached 250-fold selectivity relative to the $\sigma 1$ receptor. Also in this study, it was observed experimentally that the true hit rate improves with the docking score, similar to the study of Lyu \cite{Lyu2019}. 

\paragraph*{Kaplan et al. (2022).} Another GPCR, the 5-HT2A receptor, was targeted in an ultra-large virtual screen involving 75 million tetrahydropyridines \cite{Kaplan2022}. Several compounds with low micromolar activity were discovered. Structure-based optimization led to compounds with half-maximal effective concentration values of up to 41 nanomolar. Cryo-electron microscopy structures of the ligand-protein complexes were obtained, confirming the predicted binding modes by the docking program (DOCK) that was used in this study. 

\paragraph*{Fink et al. (2022).} The $\alpha_{2A}$ adrenergic receptor was used as the target in another ultra-large virtual screen in which 301 million compounds from the ZINC20 library were docked \cite{Fink2022}. Compounds with up to 12 nanomolar binding affinities were obtained. Also in this study, cryo-electron microscopy structures of the ligand bound to the protein confirmed the predicted binding modes of two compounds, and these structures served as the starting point for optimizing the compounds. 

\subsection{Software and Automated Routines for ULVSs}

To carry out ultra-large virtual screenings, a large number of docking procedures have to be carried out. Most docking programs only support docking a single ligand at a time, therefore virtual screening platforms or automated procedures that deploy the docking programs in parallel are required. Here, we outline two of these approaches.

\subsubsection{VirtualFlow}
\label{ssec:sw-vf}

\begin{figure}[h]
    \includegraphics[width=0.75\paperwidth]{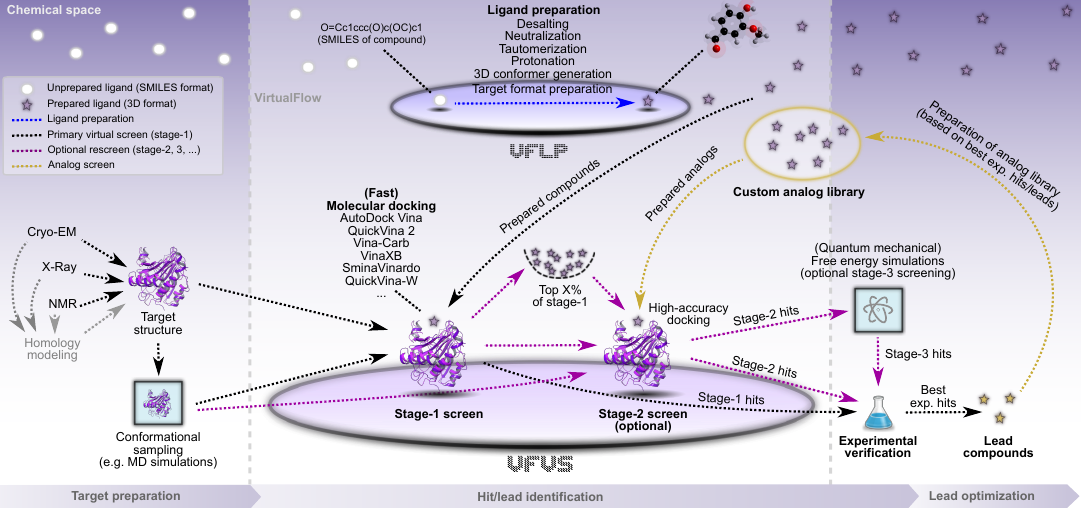}
    \caption{Overview of the VirtualFlow drug discovery platform. The first module of VirtualFlow, VFLP, can prepare large numbers of molecules from the SMILES format into a ready-to-dock format. The second module, VFVS, is able to carry out ultra-large structure-based virtual screens. Here, a variety of different docking programs can be used to dock the ligands to the target protein. The screen can be set up in a multistaged manner to improve accuracy while increasing the computation time only moderately. VFVS can also be used to optimize hit and lead compounds by searching the ultra-large ligand libraries for analogs, creating a custom analog library, and then screening this custom analog library again with VFVS.}
    \label{fig2}
\end{figure}

VirtualFlow was the first drug discovery platform dedicated to ultra-large structure-based virtual screens \cite{Gorgulla2018,Gorgulla2020a}. It is freely available, open-source, and everyone is welcome to contribute to the project (\url{https://virtual-flow.org/}). The VirtualFlow platform consists of two modules, as well as ultra-large ligand libraries in a ready-to-dock format. The two modules of VirtualFlow are VFLP (VirtualFlow for Ligand Preparation), and VFVS (VirtualFlow for Virtual Screening). An overview of the workflow of VirtualFlow is shown in Figure \ref{fig2}. VFLP and VFVS can be run using large numbers of CPUs in parallel to achieve a high ligand throughput. Both modules exhibit a perfectly linear scaling behavior with respect to the number of CPUs that are used, which was demonstrated with up to 160,000 CPUs \cite{Gorgulla2020a}. 

VFLP can generate ultra-large ligand libraries from the SMILES format into the ready-to-dock PDBQT format. The preparation procedure includes the calculation of the stereoisomers, tautomerization states, protonation states, and 3D coordinates of the molecules. 

One of the features of VFVS is that it supports a variety of external docking programs, such as AutoDock Vina \cite{Trott_2009}, Smina \cite{Koes2013}, QuickVina 2 \cite{Alhossary2015a}, QuickVina-W \cite{Hassan2017a}, Vina-Carb \cite{Nivedha}, or VinaXB \cite{Koebel2016}. Later GWO Vina and PLANTS were added \cite{Gorgulla2020b,Gorgulla2021b}. Most of these docking programs have special features, such as blind-docking capabilities (QuickVina-W), or enhanced accuracy for docking carbohydrates (Vina-Carb). VirtualFlow was used for COVID-19 drug discovery, where approximately 40 different sites on COVID-19 related proteins were targeted, resulting in over 40 billion docking instances \cite{Gorgulla2021a}. VFVS can be used in a single or multi-staged manner, where more accurate docking scenarios are used in subsequent screening steps. VFVS supports multiple ways in which the accuracy can be increased, including the increase of the conformation search by the docking programs, carrying out ensemble-dockings (to account for protein  backbone flexibility), consensus dockings, or by including side-chain flexibility via the docking programs themselves.

\subsubsection{UCSF DOCK}

Several of the success stories listed above were carried out by using the program UCSF DOCK 3.7 \cite{Coleman2013}. DOCK is a rich docking program that supports many advanced features, but was not initially designed for ultra-large virtual screenings, and therefore applying it to ultra-large ligand libraries can be more complicated than using a platform that is dedicated to ultra-large virtual screens. A detailed protocol was published by the creators of DOCK on how one can use the program to carry out ultra-large virtual screens \cite{Bender2021}. The protocol consists of over 100 steps, that not only include the virtual screening procedure, but also many other steps such as the preparation of the molecules as well as postprocessing steps.

\section{SYNTHON-BASED VIRTUAL SCREENINGS}
\label{Sec:Synthon}

Some of the ligand libraries mentioned in this manuscript are available in enumerated form, meaning that each ligand is represented explicitly in a molecular file format such as PDBQT or SMILES. An alternative representation for combinatorial libraries such, as the REAL Database, REAL Space, GalaXi Space, and CHEMriya, is to present the library implicitly via the initial fragments and the reactions that are used to combine these fragments into molecules. The fragments come along with information on how they can react with other fragments. Those fragments together with their reactivity information are called \textit{synthons}. The advantage of the synthon-based representation is that it requires dramatically less storage space than the enumerated representation mode, typically megabytes vs terabytes. One disadvantage is that the molecules in the synthon representation are not in a ready-to-dock format. 

\subsection{V-SYNTHES}

V-SYNTHES is a virtual screening method that uses the synthon representation of the REAL Space from Enamine \cite{Sadybekov2021}. In their approach, they dock first all the fragments and then assemble complete molecules out of a subset of the fragments. The selection is based on the docking scores of the fragments, and to ensure sufficient diversity several criteria are applied, such as that one reaction type cannot be part of more than 20\% of the chosen synthons. Finally, the assembled molecules are docked against the target receptor (see also Figure \ref{fig1}). The primary advantage compared to ultra-large virtual screens is that only a few million compounds are typically docked in this approach, while still exploring an ultra-large ligand library. To demonstrate the V-SYNTHES approach, in their paper the authors applied the method to the cannabinoid receptors, which led to fourteen submicromolar ligands directly out of the screen. Virtual screening-based optimization of the initial hits led to compounds with improved binding affinity, the best compounds having a binding constant of 0.9 nanomolar, which is an excellent result.

\section{DEEP LEARNING APPROACHES TO MOLECULAR DOCKING}
\label{Sec:DL}

Machine learning-based docking programs are one of the four types of docking programs, and they use a relatively large number of descriptors \cite{liu2015classification,yang2022protein,ain2015machine,li2019overview}. Like empirical docking programs, machine learning-based docking programs need training on data \cite{liu2015classification}. Even though empirical scoring functions can be considered to be machine learning-based as well (e.g. because they often use multivariate linear regression), they use parameterized functions, whereas the term machine learning-based docking programs refer to nonparametric methods that can select the final functional form \cite{ain2015machine}. Machine learning-based docking programs have a much larger number of descriptors (variables) than empirical docking methods \cite{liu2015classification}. 

The first machine learning methods for molecular docking appeared around 2004 \cite{liu2015classification}.  Deep learning is a subdivision of machine learning, and in the past few years, the first deep learning approaches have been developed for molecular docking. One advantage of deep learning approaches is that they can be orders of magnitude faster compared to classical docking programs \cite{Lu2022}. High computational speeds become crucial when processing large ligand datasets, implying that deep learning methods can be particularly beneficial for screening ultra-large libraries. Apart from considerations of speed, deep learning methods have the potential to become more accurate in the future in terms of their accuracy regarding the predicted binding affinity and the predicted docking pose than traditional docking programs. In this section, we give an overview of deep learning approaches related to molecular docking.

Docking programs predict the binding pose of a ligand to a protein, along with a score that characterizes the strength of the interaction. Historically, deep learning-based scoring functions were developed first, while the first pose-prediction methods utilizing deep learning appeared only recently. Deep learning-based scoring functions are regression models that utilize known datasets of ligand-protein interaction pairs to predict the strength of an interaction by assigning a score. New deep learning-based docking programs that include both scoring and pose prediction are still very rare. However, deep learning-based scoring functions and pose prediction methods can be combined into a full docking program, due to their modularity. Here, a deep learning-based scoring function can for instance be used to score the binding poses obtained by a deep learning-based pose-prediction method. Alternatively, a deep learning-based scoring function can be used in combination with existing docking programs to (re)rank the docking poses generated by the existing docking program by reevaluating the generated conformations by the new scoring function. A third possibility is to combine the deep learning-based scoring functions with existing sampling methods, where the scoring function is used as the cost function that is minimized during the optimization procedure. 

The new docking methods can be used in ultra-large virtual screenings, e.g. with VirtualFlow, which already supports some of the new deep learning-based methods out of the box. This can, in principle, lead to substantial speed-ups compared to the deployment of traditional docking programs in virtual screening campaigns. 

\subsection{Scoring Functions}

A wide variety of deep learning-based scoring functions have been developed in the past several years. Here in this section, we look at pure scoring functions that were published without a sampling algorithm. To our knowledge, historically one of the first such methods was  NNScore 2.0 \cite{NNScore2}. Later, a large number of other deep learning-based scoring functions have been developed. Among them are Pafnucy \cite{Pafnucy}, DeepAffinity \cite{DeepAffinity}, DeepBindRG \cite{DeepBindRG}, OnionNet \cite{OnionNet}, PotentialNet \cite{PotentialNet}, K\textsubscript{DEEP} \cite{KDEEP}, DeepAtom \cite{DeepAtom}, TopologyNet \cite{TopologyNet}, and Math-DL \cite{MathDL}, Erdas-Cicek2019 \cite{erdas2019three}, Cang2018 \cite{cang2018representability}, Atomic CNN \cite{AtomicCNN}, Francoeur2020 \cite{francoeur2020three}, and Zhu2020 \cite{zhu2020binding}. Here, as well as in the next paragraph, the name of the first author followed by the publication year was used as the name for the method in case no proper name was provided by the authors. For reviews on scoring functions, as well as for binding classification functions for protein-ligand binding see also \cite{Kimber2021,Li2021,Du2022,Crampon2022}.

\subsection{Binding Classification Functions}
Closely related to scoring functions are functions that try to distinguish ligands between binders and non-binders for a specific target, which is a classification rather than a regression task. We refer to this task in this review as binding classification. Binding classification is also referred to as \textit{virtual screening task} by some authors because classification functions are particularly useful in virtual screenings \cite{Crampon2022}. Several different classification methods based on deep learning have been developed, such as NNscore \cite{NNScore}, AtomNet \cite{AtomNet}, DeepVS \cite{DeepVS}, Ragoza2017 \cite{Ragoza2017}, DenseFS \cite{DenseFS}, Lim2019 \cite{Lim2019}, Torgn2019 \cite{Torng2019}, Tanebe2019 \cite{tanebe2019end}, Tsubaki2019 \cite{tsubaki2019compound}, Morrone2020 \cite{Morrone2020}, Sato2010 \cite{sato2010combining}, Li2019 \cite{li2019deep}, Sato2019 \cite{sato2019significance}, BindScope \cite{BindScope}, and Lim2019 \cite{lim2019predicting}. 

\subsection{Binding Pose Prediction}

The major reason why classical molecular docking methods are relatively slow is that they typically sample large numbers of ligand-protein conformations, and assign each of them a score. Deep learning methods for pose-prediction can eliminate the sampling step by directly predicting the conformation of the ligand-receptor complex.

The first pose prediction method based on deep learning might have been PoseNetDiMa \cite{PoseNetDiMa}, which uses a graph-convolution neural network and allows for flexible protein-ligand docking, where the residues at the binding site are allowed to be flexible. Another method from the same group that uses intermolecular Euclidean distance matrices (EDMs) was reported in \cite{masters2022deep}. Both methods are for local dockings (rather than blind dockings). The software of both methods is not available for download at the time of writing, but the software of the second paper will become available after publication in a peer-reviewed journal (M. A. Lill, personal communication, October 31st 2022).

DeepDock is another pioneering pose prediction model, which is based on geometric deep learning \cite{DeepDock}. However, it is rather slow according to \cite{Equibind}. EquiBind is another geometric and graph-based deep learning model, which is extremely fast with a fraction of a second per ligand using a 16-core CPU or a GPU  \cite{Equibind}. Both DeepDock and EquiBind are freely available. TANKBind has tried to improve upon the results of EquiBind \cite{TANKBind}. TANKBind improves on the accuracy of EquiBind by using trigonometric constraints in the model, and by dividing the protein into multiple regions.

Both EquiBind and TANKBind are substantially faster than traditional docking programs. DiffDock, which is a diffusion generative model, was developed by the same research group as EquiBind, and their results indicate that it can be more accurate than any of the two previously mentioned blind docking methods \cite{DiffDock}. On the other hand it is considerably slower than EquiBind or TANKBind, however, it is still faster than most traditional docking programs. 

Most of the above pose prediction methods carry out blind pose prediction, meaning that the entire protein surface is considered as potential binding area, rather than a restricted binding site.

\subsection{Full Dockings}

Complete docking programs using deep learning, meaning for both pose prediction as well as scoring, are still very rare as most approaches so far use deep learning either on pose prediction or on scoring. GNINA \cite{Gnina_17,ragoza2017ligand,hochuli2018visualizing,sunseri2021virtual,sunseri2019convolutional,mcnutt2021gnina}, is a fork SMINA \cite{Koes2013}, and SMINA itself is a fork of AutoDock Vina \cite{Trott_2009}. GNINA was one of the first full docking programs using deep learning. It uses convolutional neural networks (CNNs) for the scoring function and uses a traditional sampling algorithm to explore the conformational space. Another docking program that is conceptually similar to GNINA in that it uses CNNs in combination with traditional search algorithms is MedusaDock \cite{MedusaDock} in combination with MedusaNet \cite{MedusaNet}. MedusaNet is a CNN-based scoring function, and MedusaDock provides the sampling algorithm. MedusaGraph is another docking method of the same research group, which is based on graph neural networks \cite{MedusaGraph}. In contrast to MedusaDock-MedusaNet, it can directly predict the docking pose without extensive sampling using traditional methods. MedusaGraph is 10-100 times faster than traditional docking programs. However, it requires an initial docking pose, which can be generated for instance with MedusaDock, which subsequently reduces the aforementioned speedup. MedusaGraph uses the scoring function MedusaScore which is part of MedusaDock. MedusaNet and MedusaGraph are freely available. TANKBind, which was mentioned in the previous subsection, is a full docking program as well \cite{TANKBind}.

\section{MACHINE LEARNING APPROACHES TO VIRTUAL SCREENINGS}
\label{Sec:MLVS}

Deep learning-based docking programs are one way to use deep learning in structure-based virtual screenings to reduce computational costs and potentially improve the accuracy of the dockings. Another way to use deep learning, and more generally machine learning in virtual screenings is to apply it on the screening level rather than only on the docking level. Here, ligand-based machine learning models are usually trained by carrying out molecular docking on a small subset of compounds from the entire ultra-large ligand library. This elegant combination of LBVSs and SBVSs in ML-LB-ULVS can reduce computational costs dramatically. Subsequently, the entire library is screened with the trained ligand-based ML model (see also Figure \ref{fig1}). We outline several of these deep learning approaches below, which focus on ultra-large structure-based libraries. 

\subsection{Deep Docking}

Deep Docking was one of the first deep learning approaches for ultra-large virtual screenings \cite{DeepDocking}. Here, a small number of molecules (typically around 1\% of the entire library) is docked, and then a quantitative structure-activity relationship (QSAR)-based deep learning model is trained on them in an iterative fashion. The majority of the ligand library is then screened using the fast QSAR model, resulting in a speed-up of around 100 compared to the docking of the entire library. The authors have applied their method by screening over one billion compounds against twelve drug targets. In a separate paper, the authors applied Deep Docking to SARS-CoV-2 drug discovery, where they targeted the protein M\textsubscript{pro} with approximately 40 billion molecules \cite{DeepDockingCovid}. The code of Deep Docking is freely available, and a detailed protocol of the method is discussed in \cite{DeepDockingProtocol}.

\begin{marginnote}[]
    \entry{Quantitative structure-activity relationship (QSAR)}{A regression or classification model that relates structural features of molecules to their biological activities.}
\end{marginnote}

\subsection{Active Learning Workflow with AutoQSAR/DeepChem (AQ/DC)}

Similar to the Deep Docking approach, in the AutoQSAR/DeepChem (AQ/DC) approach, a relatively small number of ligands of the library is used for training a deep learning QSAR model in addition to molecular fingerprints \cite{AQDC}. The QSAR model uses graph-convolutional neural networks via the DeepChem package. Optionally active learning can be used to refine the model during the workflow. Around 5\% of the top hits are then docked with regular docking programs, resulting in an overall speedup of around 15-20 compared to the standard ultra-large virtual screenings. In the same paper, the authors applied their new method to the target proteins AmpC and the D4 dopamine receptor, which they had targeted previously with standard ultra-large virtual screens \cite{AQDC,Lyu2019}. This resulted in a recovery of approximately 80\% of the previously experimentally confirmed hit compounds. 

\begin{marginnote}[]
    \entry{Molecular fingerprint}{A binary representation of a molecule, where each bit represents a certain structural property.}
\end{marginnote}

\subsection{MolPAL}

Another deep learning-based approach is MolPAL \cite{MolPal}, which stands for \textit{Molecular Pool-based Active Learning}. MolPAL uses Bayesian optimization as well as molecular fingerprints and was tested with surrogate models based on random forest, feedforward neural networks, and directed-message passing neural networks. Only a small fraction of molecules is docked (around 2.5\%), i.e. ones more likely to be scored favorably, resulting in a speedup of approximately 40. In their study, roughly 90-95\% of the top 50,000 hits in a virtual screening of around 100 million compounds are recovered. MolPAL is freely available and open-source.

\section{HARDWARE ACCELERATION OF MOLECULAR DOCKINGS WITH GPUS}
\label{Sec:GPU}

Virtual screenings, in particular at the ultra-large scale, typically require a large amount of computation time \cite{Gorgulla2020a}. Therefore, often computer clusters or the cloud are needed to carry them out. Almost all traditional docking programs run on CPUs, and typically require seconds to minutes per ligand. In the previous sections, we have seen multiple ways on how the runtime can be reduced, by effectively reducing the number of traditional dockings required via the synthon-based approaches, by combining ligand-based virtual screens with structure-based screens to give rise to ML-LB-ULVSs, or by using deep learning-based docking methods. Another approach, which can be used in combination with the above approaches, is to use graphics processing units (GPUs) instead of CPUs. GPUs can have up to 10-100 times better price/performance ratio than CPUs.

Docking programs for GPUs have appeared only in the past few years. Some of them try to port traditional docking programs to GPUs, such as AutoDock-GPU \cite{santos2021accelerating}, Vina-GPU \cite{VinaGPU}, Uni-Dock \cite{UniDock}, and MedusaDock GPU \cite{MedusaDockGPU}, while others are new docking programs tailored for GPUs (e.g. Accelerated CDOCKER \cite{CDOCKERGPU}). In addition, most of the deep learning-based docking programs that were discussed above run on CPUs and GPUs, e.g. GNINA \cite{mcnutt2021gnina}, MedusaDock-MedusaNet \cite{MedusaNet}, MedusaGraph \cite{MedusaGraph}, and TANKBind \cite{TANKBind}. 

\paragraph*{AutoDock-GPU.} AutoDock-GPU \cite{santos2021accelerating} is the GPU version of the well-known docking program AutoDock 4 \cite{AutoDock4}. AutoDock-GPU is available both as OpenCL as well as CUDA version and can run on CPU, GPU, and FPGA architectures \cite{AutoDockGPU_FPGA,AUTODOCKGPUOPENCL}. Compared to the single-threaded CPU version, the speedup is up to a factor of around 350 when using GPUs. However, since AutoDock 4 was very slow in general, the GPU-accelerated version is roughly as fast as more modern CPU-based docking programs \cite{UniDock}. The authors applied the new software to multiple case studies, including the docking of approximately 1 billion molecules for COVID-19 drug discovery in less than 24 hours on the Summit supercomputer.

\paragraph*{Vina-GPU.} AutoDock Vina \cite{Trott_2009} is the most cited docking program of all time, with over 20,000 citations at the time of writing. In addition, several benchmark studies show that it is among the most accurate docking programs. A GPU version of AutoDock Vina is therefore of great interest. One attempt was the software Viking, however, no speedup has been reported \cite{VinaGPU}. The first successful GPU version of AutoDock Vina with a significant speedup, Vina-GPU, was published in 2022 \cite{VinaGPU}. It makes minor modifications to the algorithm of AutoDock Vina, but the final docking scores are almost identical. Vina-GPU exhibited a speedup between 21 and 50 when using an NVIDIA RTX-3090 compared to AutoDock Vina running on a 10-core Intel\textsuperscript{\textregistered} Core\textsuperscript{TM} i9-10900K CPU at 3.7 GHz. 

\paragraph*{Uni-Dock.} Another software building on AutoDock Vina is Uni-Dock \cite{UniDock}. In contrast to Vina-GPU, it builds on AutoDock Vina 1.2 \cite{Vina1_2}, while Vina-GPU is based on the original AutoDock Vina. Uni-Dock claims to have greater acceleration than Vina-GPU, and it also supports multiple scoring functions: Vina, Vinardo, and the AutoDock 4 scoring functions. Uni-Dock achieves a speedup of over 1000 when running on an NVIDIA V100 32G GPU compared to AutoDock Vina using a single core of an Intel\textsuperscript{\textregistered} Xeon\textsuperscript{\textregistered} Platinum 8269CY (Cascade Lake) CPU running at 2.5 GHz. Uni-Dock is not freely available and executable, but available via a webserver (\url{https://academic.dp.tech/toolkits/uni-dock-serving/}). 

\paragraph*{MedusaDock GPU.} MedusaDock GPU \cite{MedusaDockGPU} is a GPU-enabled version of MedusaDock \cite{MedusaDock}. The speedups observed are moderate, in the lower single-digit range, when comparing MedusaDock running on a single core of an Intel\textsuperscript{\textregistered} Xeon\textsuperscript{\textregistered} E5-2680 v4 (Broadwell) and MedusaDock GPU running on an NVIDIA TESLA P100 GPU.

\paragraph*{Accelerated CDOCKER.} CDOCKER (short for \textit{CHARMM Docker}) and flexible CDOCKER are docking methods that are part of the CHARMM package \cite{CDOCKER,CDOCKER_FLEX}. Accelerated CDOCKER is a GPU-accelerated version of the CDOCKER and flexible CDOCKER methods \cite{CDOCKERGPU}. It uses the fast Fourier transform as a key algorithm for the GPU acceleration and achieves an impressive speedup of approximately 15,000 when using an NVIDIA GeForce GTX 1080 vs the original CDOCKER running on an Intel\textsuperscript{\textregistered} Xeon\textsuperscript{\textregistered} Processor E5645 at 2.4 GHz. CDOCKER also includes a parallel MD-based simulated annealing method that can run on GPUs in parallel, with an observed speedup of approximately 20.
 
\paragraph*{Deep learning-based docking methods.} Most of the deep learning-based docking programs can use GPUs since deep learning models can usually be run on GPUs for their evaluation. Among them are GNINA, MedusaDock-MedusaNet, MedusaGraph, and TANKBind. In addition, most of the pure scoring functions, classification functions and pose prediction methods based on deep learning can be run on GPUs. However, they would need to be implemented as full docking methods to be useful for most virtual screening campaigns.

\section{HIGH-THROUGHPUT FREE ENERGY SIMULATIONS}
\label{Sec:HTFES}

So far in this review, we have seen several examples of how multiple virtual screening stages can be used to save computation time, e.g. by using a very fast virtual screen (such as ML-LB-ULVSs or deep learning based docking methods in standard ULVS) of the entire library as the first stage. Such virtual screens are generally less accurate than full molecular dockings. Staging can also be used in the opposite direction, to increase the accuracy compared to molecular dockings. Here, free energy simulations based on molecular dynamics (MD) simulations can be used to improve the accuracy of the binding affinity estimates for the best compounds of the previous stage. Free energy simulations are in theory exact (relative to the force field used), provided that the free energy converges (for that the simulations need to run long enough). They take into account the full dynamics of the molecular system, can treat all solvent molecules explicitly, and include all enthalpic and entropic components of the free energy. When applied to a larger number of systems in an automated way, free energy simulations are also referred to as high-throughput free energy simulations (HTFESs).

In addition, quantum mechanics/molecular mechanics (QM/MM) methods can be used to treat the parts of the system (typically the ligand, the binding site of the protein, and possibly some solvent molecules at the binding site) with higher accuracy \cite{Gorgulla2018}. This can for example be beneficial if metals are located at the binding site of the protein or part of the ligand. 

Hardware acceleration, in particular the usage of GPUs, is key regarding HTFESs. Multiple molecular dynamics software packages have been optimized for GPUs in the past 15 years, such as GROMACS \cite{abraham2015gromacs}, AMBER \cite{Amber2022}, CHARMM \cite{brooks2009charmm}, and NAMD \cite{NAMD}, which greatly enhances their simulation speed. In principle, these MD simulation packages can be directly used to carry out free energy simulations of protein-ligand systems. However,  applying them to a large number of ligands will likely require custom scripting to automate certain processes. In the past few years, several high-throughput MD and free energy simulation methods have been developed, which automate many processes and allow to handle large numbers of protein-ligand systems.

\begin{marginnote}[]
    \entry{Free energy methods}{Methods for predicting the free energy differences between different states of a molecular system. They can be used to predict the binding affinity between molecules.}
    \entry{Molecular dynamics (MD) simulation}{Computer simulation of a molecular system in which the movements of each atom are calculated for many time steps.}
    \entry{Quantum mechanics/molecular mechanics (QM/MM) method}{Molecular models that treat one part of the molecular system on the quantum level and the rest using classical mechanics.}
\end{marginnote}

\paragraph*{HTMD.} One of the first packages dedicated to high-throughput molecular dynamics was HTMD \cite{doerr2016htmd}. It supports a large number of different tasks and features, such as molecular dynamics, Markov state modeling, clustering, free energy calculations, and projection methods, adaptive sampling, and can run in the cloud. The package has been extended to also allow simulation of membrane proteins in an automated way, and it is freely available for academic groups \cite{HTMD_membrane}. 

\paragraph*{BRIDGE.} A package that is dedicated to high-throughput MD simulations that support free energy calculations is BRIDGE (Biomolecular Reaction and Interaction Dynamics Global) \cite{senapathi2020bridge}. It is an open-source web platform and built on top of the Galaxy platform. BRIDGE supports absolute and relative binding free energy simulations. It also allows to easily share obtained simulation data and supports independent verification of the data by external people, which can for instance be useful for other scientists or during the peer review process. 

\paragraph*{Deep learning-based force fields.} Traditionally, classical force fields such as AMBER and CHARMM have mostly been used in MD simulations and thus free energy simulations. Recently, deep learning-based force fields have been developed, such as ANI (Accurate neural network engine for molecular energies) and AIMNet (atoms-in-molecules net), which have increased the accuracy of absolute binding free energy simulations by up to 50\% \cite{ANI,AIMNet}. 
\begin{marginnote}[]
\entry{Force field}{A method that predicts or specifies the forces that act between atoms in a molecular system.}
\end{marginnote}

\paragraph*{Case study with OpenMM.} High-throughput free energy simulations to estimate the ligand-protein binding affinity have also been carried out on a larger scale in \cite{guterres2020improving}. Here, the authors have used AutoDock Vina to dock the ligands to the target proteins to obtain an initial docking pose, and then run free energy simulations using OpenMM. 56 target proteins were used, and 560 ligands in total. The results show that the estimated binding free energies have improved in accuracy compared to the original AutoDock Vina scores. The area under the ROC curve improved by 22\%.

\paragraph*{Quantum mechanical simulations.} Packages for quantum mechanical simulations suitable for free energy calculations of protein-ligand systems exist as well. In principle, many of the standard packages for MD simulations provide some form of QM/MM support, such as CHARMM \cite{brooks2009charmm} and AMBER \cite{Amber2022}. However, they are mostly focused on classical simulations. TeraChem \cite{seritan2021terachem} was the first package for quantum mechanical simulations specifically written for GPUs, resulting in a remarkably high performance. It can simulate entire proteins on the density functional theory level, and can also carry out QM/MM simulations. Another package that is currently developed for exascale supercomputers is NWChemEx \cite{NWChemEx}, which will support quantum mechanical simulations of molecular systems considerably larger than possible today.

\section{CONCLUSIONS AND OUTLOOK}
\label{Sec:Outlook}

Structure-based virtual screenings have evolved into a powerful tool for drug discovery in the past few years. Ultra-large virtual screens allow to access billions of compounds, allowing to find exceptionally tight-fitting ligands for even highly challenging target sites. This approach allows to drug  new target classes such as allosteric sites or protein-protein interactions, which with approximately 500,000 types are considered to be one of the holy grails of drug discovery. Machine learning and deep learning techniques on the molecular docking and virtual screening levels allow to speed up the calculations by orders of magnitude, as do other algorithmic approaches such as synthon-based methods, or the use of hardware accelerators such as GPUs. It is expected that virtual screenings will further expand the accessed chemical space by orders of magnitude in the coming years, with no limits in sight given the vastness of the chemical space of molecules suitable for drug discovery. Deep learning approaches for molecular docking and for free energy simulations have the potential to significantly improve the accuracy and reliability of free energy predictions, though significant works lies still ahead to achieve this goal. GPUs and other hardware accelerators will aid in accessing larger regions of the chemical space of druglike molecules in the future. GPUs will be key for achieving highly accurate free energy simulations, whether on the classical or the quantum level, which hold great promise to improve the reliability of binding affinity prediction. In addition, structure-based virtual screens can be extended to tasks other than hit and lead discovery and optimization, such as predicting off-target binding via inverse virtual screenings or combining them with the prediction of pharmacokinetic properties \cite{Gorgulla2022}.

In summary, structure-based virtual screens are on the one hand already a mature technique, and able to be routinely applied in drug discovery campaigns to identify potent hit and lead compounds and to optimize them. On the other hand, they have only recently become powerful enough to be of real value in actual drug discovery projects, and have barely begun to scratch the surface of what they can achieve in the future. A bright future therefore likely lies ahead for structure-based virtual screens.

\begin{summary}[SUMMARY POINTS]
    \begin{enumerate}
        \item Early-stage drug discovery using traditional methods such as experimental high-throughput screens faces several major problems, such as high costs, labor, and poor initial hit compounds. Structure-based virtual screenings can solve most of these problems, by reducing the costs and workload, and by improving the quality of hit as well as lead compounds.
        \item Ultra-large structure-based virtual screenings in particular have proven to be able to discover highly potent hit and lead compounds, as well as to optimize initial hit compounds. This in turn leads to the ability to find potent binders to more challenging target classes such as protein-protein interactions.
        \item Synthon-based virtual screenings allow exploration of ultra-large ligand libraries without the need to screen the entire library, but only a small fraction of it. This allows reducing the computational costs by a factor of around 100. 
        \item Deep learning methods for molecular dockings can be computationally much faster than traditional docking methods, and can in some cases already exceed their accuracy. These methods can be applied during ultra-large virtual screens as well as in synthon-based approaches.
        \item Machine learning approaches can be applied on the virtual screening level to dramatically speed up the computations, which is particularly beneficial when screening ultra-large libraries. Here, normally ligand-based machine learning models are trained on a training set obtained by applying structure-based virtual screens on a smaller part of the library and then applying the trained model on the entire ligand library. 
        \item GPUs can be used instead of, or in complement with, CPUs. They can provide one to two orders of magnitude better cost/performance ratio than CPUs. Several docking programs are available that can make use of GPUs, such as AutoDock-GPU, Vina-GPU, and most of the deep learning-based docking methods. 
        \item Free energy simulation methods to estimate the binding affinity of ligands to receptors can be more accurate than molecular dockings. They can model the complete flexibility of the receptor, as well as solvent molecules. High-throughput free energy simulations have been devised that can be employed as an additional stage in structure-based virtual screening approaches.
        \item There have been multiple successive stories involving ultra-large virtual screens, which included experimental validation. In these studies, nanomolar to picomolar binders have been reported for different types of receptors, including enzymes, GPCRs, and protein-protein interactions. 
\end{enumerate}
\end{summary}

\begin{issues}[FUTURE ISSUES]
    \begin{enumerate}
        \item The chemical space of small druglike molecules is estimated to contain over $10^{60}$ molecules, and even ultra-large virtual screens only sample a tiny fraction of this space. Further extension of structure-based virtual screens to larger parts of the entire chemical space of druglike molecules are promising. 
        \item Deep learning approaches to molecular docking have the potential to significantly improve their accuracy. This is of high importance, as current molecular docking programs are highly inaccurate regarding their binding affinity estimates. Large, high-quality datasets will be key to achieving this goal with deep learning methods. 
        \item Machine learning approaches to virtual screens hold great promise to enable the exploration of larger regions of the chemical space than possible today. The same is true of new methods that reduce the number of molecules that need to be explicitly screened, and intelligently decide which molecules are promising for a particular target protein. 
        \item Further utilization of hardware acceleration via GPUs and other hardware (such as Google's Tensor Processing Units) has the potential to further reduce the costs of virtual screens while allowing for large chemical spaces to be explored. 
        \item High-throughput free energy simulations have great potential to improve the accuracy in structure-based virtual screens via multi-staging approaches. To date, only very few tools are available to aid with high-throughput free energy simulations, and more convenient and easy-to-use tools are needed. Ideally, these can be applied in an automatic way after docking-based virtual screening stages.
        \item Quantum-mechanical scoring functions for molecular docking could improve the accuracy, in particular when there are metals at the active site. In the long term, quantum computers might provide substantial speed-ups for quantum mechanical scoring functions for protein-ligand docking. 
        \item While there exist already a few web servers for virtual screens, they are mostly not yet available for ultra-large virtual screens. Visual interfaces that interface with the cloud for ultra-large virtual screens that are easy to use would facilitate the wider adoption of structure-based virtual screens by the drug development communities. 
        \item Structure-based virtual screens could be of great value to predict off-target effects, for instance via inverse virtual screens of specific small molecules against the entire human proteome. The AlphaFold Structure Database essentially opened the door for such approaches. This could reduce the risk of toxicity of new drug candidates. 
    \end{enumerate}
\end{issues}

\section*{DISCLOSURE STATEMENT}
C.G. is a cofounder of Quantum Therapeutics Inc., a company that uses
computational methods for drug development. C.G. is a cofounder of Virtual Discovery Inc., which is a fee-for-service company for computational drug discovery.

\section*{ACKNOWLEDGMENTS}
C.G. would like to thank Brigitte Klein, Konstantin Fackeldey, and Akshat Nigam for proofreading the manuscript and for their valuable feedback. C.G. would like thank ChemAxon for a free academic license for their software. 

\bibliographystyle{ar-style3.bst}
\bibliography{Manuscript}

\end{document}